# Toward Practical-Scale Quantum Annealing Machine for Prime Factoring

Masaaki Maezawa[1]*, Go Fujii[1], Mutsuo Hidaka[1], Kentaro Imafuku[1], Katsuya Kikuchi[1], Hanpei Koike[1], Kazumasa Makise[1], Shuichi Nagasawa[1], Hiroshi Nakagawa[1], Masahiro Ukibe[1], and Shiro Kawabata[1]

[1]*National Institute of Advanced Industrial Science and Technology, Tsukuba, Ibaraki 305-8565, Japan*

We propose a prime factorizer operated in a frame work of quantum annealing (QA). The idea is inverse operation of a multiplier implemented with QA-based Boolean logic circuits. We designed the QA machine on an application-specific-annealing-computing architecture which efficiently increases available hardware budgets at the cost of restricted functionality. The invertible operation of QA logic gates consisting of superconducting flux qubits was confirmed by circuit simulation with classical noise sources. The circuits were implemented and fabricated by using superconducting integrated circuit technologies with Nb/AlO$_x$/Nb Josephson junctions. We also propose a 2.5-dimensional packaging scheme of a qubit-chip / interposer / package-substrate structure for realizing practically large-scale QA systems.

## 1. Introduction

Since Shor's discovery of quantum algorithms,[1] prime factoring of large numbers has been the biggest game in the field of quantum computing[2-4] because its difficulty guarantees the security of RSA cryptosystem.[5] The first experiment of prime factoring by Shor's algorithm factored 15 into 3 and 5 using the liquid nuclear magnetic resonance (NMR) technique.[6] Afterward, several experiments using different methods, *e.g.*, quantum optics[7] and silicon photonics,[8] followed and demonstrated "15 = 3 × 5". On the other hand, a new computing concept based on quantum mechanics, *i.e.*, quantum annealing (QA) was proposed theoretically and examined numerically by Kadowaki and Nishimori.[9] A hardware implementation of the QA concept, a QA machine, has been developed and eventually commercialized by D-Wave Systems in the past decade.[10-15] While D-Wave-type QA machines are considered promising for solving optimization and sampling problems, they can also solve various problems expressed in forms of combinational logic on the basis of

Ising-model implementation of Boolean logic.[13)] In this paper we present design of a superconducting QA-based factorizer which is basically the same as a Boolean multiplier operated in the inverse direction. Our challenge toward a practical-scale factorizer is discussed from concept to technology.

## 2. Quantum Annealing for Factoring

The QA process is basically invertible because of crawling-on-ground-state operation which starts and finishes with the minimum energy state. Figure 1 shows an operation principle of a factorizer that is a QA multiplier operated inversely.[16)] The QA machine, multiplier and/or factorizer, is designed to have the minimum energy when the boundary states satisfy $P = M \times N$. By fixing the factors $M$ and $N$, the product $P = M \times N$ is obtained at the energy minimum after the annealing process. In a similar way, by fixing the product $P$, one of the possible combinations of the factors $M$ and $N$ is probabilistically obtained. The factoring operation is not reversible because information entropy reduces during the multiplication operation. For example, for factoring of $P = 15$, the QA factorizer gives one of the four possible answers $(M, N) = (1, 15), (3, 5), (5, 3)$ or $(15, 1)$. While the uncertainty is a disadvantage for some applications, the QA-based factoring is still applicable to our main target, RSA cryptanalysis, for which a number $P$ to be factorized is always a product of two prime numbers $M$ and $N$.

The QA factoring machine is simply designed as a multiplier composed of Boolean logic gates in the same way as conventional digital circuits. But differently from the digital circuits, the logic gates consist of qubits and inter-qubit couplers to be operated in a scheme of annealing. The QA implementation of the Boolean logic gates is expressed by Hamiltonian of Ising model,

$$H = \sum_{i} h_i \sigma_i + \sum_{i,j} J_{ij} \sigma_i \sigma_j , \qquad (1)$$

where, $\sigma_i$ is an $i$-th spin, $h_i$ is a bias energy for the $i$-th spin, and $J_{ij}$ is a coupling energy between the $i$-th and $j$-th spins. For instance, a QA NOR gate has $h_1 = h_2 = 0.5$, $h_3 = 1$, $J_{12} = 0.5$ and $J_{13} = J_{23} = 1$ (Fig. 2(a)). For these energy coefficients, eq. (1) yields the minimum energy of $H = -1.5$ when the states of $Q_1$, $Q_2$ and $Q_3$ satisfy the correct NOR relation, *i.e.*, $(\sigma_1, \sigma_2, \sigma_3) = (-1, -1, 1), (-1, 1, -1), (1, -1, -1)$ or $(1, 1, -1)$. No other combination can yields a smaller energy of $H < -1.5$.

An arbitrary Boolean logic circuit is constructed by accordingly connecting the QA logic gates with $J_{ij} = -1$. As shown in Fig. 2(b), a QA half adder, which is one of the chief components in a conventional implementation of the multiplier, is composed of three QA NOR gates, $Q_1$-$Q_2$-$Q_3$, $Q_4$-$Q_5$-$Q_6$ and $Q_7$-$Q_8$-$Q_9$. Here, the inter-gate couplings with $J_{ij} = -1$, *e.g.*, the $Q_3$-$Q_8$ coupling with $J_{38} = -1$, plainly interconnect the qubits whereas the inter-gate couplings with $J_{ij} = 1$, *e.g.*, the $Q_1$-$Q_4$ coupling with $J_{14} = 1$, provide not only interconnection but also NOT operation. Note that the NOR function is universal in Boolean logic.

While a Boolean multiplier can be constructed by using the QA half adders in Fig. 2(b) and QA AND gates synthesized by changing the sign of $h_1$, $h_2$, $J_{13}$, and $J_{23}$ in the NOR Hamiltonian, more efficient logic synthesis is possible as shown in Fig. 2(c) where a unit cell, *i.e.*, a 1 x 1 multiplier is comprised of 6 qubits. In the same way as the QA NOR, Hamiltonian of the QA multiplier unit is designed to have the minimum energy when the states of the qubits satisfy the multiplication relation, that is, $\sigma_5$ and $\sigma_6$ are respectively a carry and sum of $\sigma_1 \cdot \sigma_2 + \sigma_3 + \sigma_4$. A QA factorizer is simply implemented by arranging a two dimensional network of $m \times m$ QA multiplier units, which can factorize an integer up to $P_{max}$, where $P_{max} = 2^m$. If we do not care about trivial solutions such as $P = P_{max} \times 1$, $P = P_{max}/2 \times 2$, $P = P_{max}/3 \times 3$, and so on, the $m \times m$ circuit is applicable to a factoring problem for an integer larger than $P_{max}$.

## 3. Architecture and Implementation

The QA machine for practical-scale factoring is to be realized by using superconducting integrated circuit technology, the most advanced among solid-state quantum circuit technologies.[17,18] However, a strong constraint in the architecture layer is the limited hardware budget associated with the immaturity of the superconducting circuit manufacturing. Even state-of-the-art Nb-based technologies allow the integration of only $10^4$ to $10^5$ Josephson junctions per chip,[19,20] which is four to five orders of magnitude lower than transistor counts in today's semiconductor technology. Although the superconducting circuit technology is steadily progressing as plotted in Fig. 3, it is too optimistic to expect that it will reach a comparable level to the semiconductor technology in the near future. Our strategy against the hardware restriction is restriction of the functionality: we intend to develop a QA machine only for factoring applications. The target QA machine has a fixed structure of the qubit network with built-in parameters, $h_i$ and $J_{ij}$, specialized for the factoring operation. The special-purpose approach, which we name an Application Specific Annealing Computing (ASAC) architecture, efficiently reduces the hardware overhead as well as the cost and time for development. Although the limited faculty limits the applications, we expect a complementary role with general purpose QA systems like D-Wave machine like the coexistence of general-purpose FPGA (field-programmable gate array) and special-purpose ASIC (application specific integrated circuit) in the semiconductor industry.

The circuits are implemented with Josephson junctions, superconducting inductors, and flux transformers. Following D-Wave,[11-14] we employ a superconducting flux qubit whose potential is tunable with transverse fields. In contrast with D-Wave machine, the inter-qubit coupler is a passive transformer with fixed mutual inductance. Figure 4 shows a simplified schematic of our qubit consisting of a main inductance loop including $L_q$ and $L_x$, and a 2-Josephson-junction SQUID (superconducting quantum interference device) with small inductors of $L_t$, which is basically the same as QFP (quantum flux parametron) invented by E. Goto more than thirty years ago.[30,31] The state of the flux qubit, "0" or "1", is defined by the direction of a qubit circulating current $I_q$: we define the "0" state as $I_q$ flowing in the counterclockwise direction and the "1" state as $I_q$ flowing in the clockwise direction. The qubit bias $h_i$ is set with a current $I_x$ which applies a magnetic flux $\Phi_x = I_x M_x$ to the main loop of the qubit. The fixed coupling strength $J_{ij}$ is proportional to the effective mutual inductance $M_{ij}$ between the qubits. The transverse field for annealing is controlled by an external control current $I_t$ which induces a magnetic flux $\Phi_t = I_t M_t$ in the 2-Josephson-junction SQUID.

## 4. Simulation with Classical Noise

The QA NOR circuits were simulated with classical noise sources in order to confirm the concept of the inverse logic operation of the annealing circuits. The stochasticity for annealing was emulated with classical current noise sources connected to the Josephson junctions in parallel. For the simulation, the Josephson junctions with $I_c = 4$ µA were also shunted with resistance of 3.2 kΩ and capacitance of 17 fF, assuming a Nb/AlO$_x$/Nb Josephson junction technology with a Josephson critical current density $J_c$ of 1 µA/µm$^2$. The Josephson junction parameters yield McCumber-Stewart parameter of 10$^4$. The current noise was programmed to have a Gaussian amplitude distribution with a standard deviation of 0.13 µA and a sampling frequency of 2 THz, which approximates Johnson-Nyquist noise of the 3.2-kΩ shunt resistance at 1 K within a bandwidth of 1 THz.

Figure 5 shows example results of inverting operation of the QA NOR gate. An extra qubit $Q_4$ with a slightly stronger bias of $h_4 = -1.1$ or $+1.1$ was added to control the state of the output qubit $Q_3$ (Fig. 5(a)). Figure 5(b) is a schematic for the simulation consisting of the flux qubit in Fig. 4. Here, for simplicity, the 3.2-kΩ resistor, 17-fF capacitor, and 1-K thermal noise source are not shown. While the qubits have different values of the coupling mutual inductance $M_{ij}$, the total inductance of the qubit main loop is maintained to be 260 pH for all the qubits. The states of $Q_i$ are defined to be "0" when $I_{qi}$ flows in the counterclockwise direction and to be "1" when $I_{qi}$ flows in the clockwise direction, where $i$ = 1, 2, 3, and 4.

The resulting simulated waveforms show correct inverse operations of the NOR gate (Fig. 5(c)). The high and low levels in the $I_t$ waveform correspond to the zero and maximum transverse fields, respectively. In the $I_{qi}$ waveforms, the low level corresponding to the flow in the counterclockwise direction represents the "0" state and the high level corresponding to the flow in the clockwise direction represents the "1" state. When the negatively over-biased $Q_4$ fixed the $Q_3$ state at "0" with $\sigma_3 = -1$, the states of $Q_1$ and $Q_2$ were one of the correct combinations of NOR, $(\sigma_1, \sigma_2) = (-1, 1)$, $(1, -1)$ or $(1, 1)$. When the positively over-biased $Q_4$ fixed the $Q_3$ state at "1" with $\sigma_3 = 1$, both of $Q_1$ and $Q_2$ was in "0" state, $(\sigma_1, \sigma_2) = (-1, -1)$, which is also the correct NOR operation. Table I summarizes results of 200 shots of the inverse NOR annealing, which includes no error operation. The obtained results, though taking account of no quantum effect, indicate that the invertible logic operation is highly feasible in QA systems. Besides, the results also suggest the possibility of a classical annealing machine which might be slower than the QA machine but could operate at higher temperature.[32]

## 5. Technology Integration

As discussed at the beginning of §3, the number of available Josephson junctions or superconducting qubits per chip is strictly limited. At present, for instance, the existing largest-scale QA machine, D-Wave 2000Q, has at most 2048 qubits, which consists of a single QA processor chip.[10] Multi-chip module (MCM) packaging is therefore an essential technology for realizing a practically large-scale QA machine for solving practical problems. Another big challenge in the technology layer is how to meet the different technology requirements for different types, quantum and classical, of circuits. On one hand, low-$J_c$ Josephson junctions in submicron sizes are necessary for the quantum circuits, qubits and couplers, to reduce both subgap leakage currents and parasitic capacitance for achieving long decoherence times. On the other hand, higher-$J_c$ junctions in moderate sizes are suitable for fabricating the classical circuits for control and readout of the qubits. Our decision is separation of the fabrication processes: the qubits and couplers are integrated on reasonably small-size chips for maintaining sufficient yields of working chips; the auxiliary control and readout circuits are built in large-size interposers for facilitating the MCM packaging.

Figure 6 shows a schematic cross-section of a 2.5-dimensional packaging, a QUbit-chip /

Interposer / Package-substrate (QUIP) structure, which we propose for realizing practical-scale QA factorizers on the ASAC architecture. A qubit chip on which the qubits and couplers are densely integrated is fabricated by using a modified version of our standard technology developed chiefly for superconducting digital circuits.[18] The modification includes the deposition of Nb/AlO$_x$/Nb Josephson-junction trilayers directly on hydrogen-terminated undoped Si wafer and the elimination of noisy disordered oxide such as anodized NbO$_x$, which are expected to improve the qubit performance. Figure 7 schematically shows a cross-section of the qubit chips including four Nb layers M1, M2, M3 and M4, planarized SiO$_2$ insulation layers, and a normal resistance layer R. Josephson junctions with $J_c$ = 1 µA/µm$^2$ are fabricated with a trilayer structure consisting of a 200-nm-thick Nb base electrode M1, a 20-nm-thick Al/AlO$_x$ barrier, and a 100-nm-thick Nb counter electrode JJ. An active interposer, on which the qubit chips are mounted by using a flip-chip technology with superconductor bumps, includes the control and readout circuits consisting of moderate-size Josephson junctions. Superconducting through silicon vias (TSVs) are also fabricated in the active interposers for package interconnection (Fig. 8). A package substrate, on which the active interposers with the qubit chips are packaged, is used for arranging a large number of wirings to connect the QA machine and room temperature electronics. In addition, bridge interposers are introduced for connecting the adjacent active interposers on the package substrate. One of the key technologies for QUIP packaging is a flip-chip bonding technique with superconductor bumps to enable superconducting, *i.e.*, quantum interconnection between the qubit chips via the interposer. Toward quantum chip-to-chip communication we are developing a preparation technology for high-density superconductor bumps made from a lead alloy (Fig. 9).

Another issue to be considered is the substrate material of the QUIP components. To avoid the collapse of the QUIP-packaged system caused by thermal contraction during cooling, all the components, the qubit chips, active and passive interposers, and package substrates, should have similar or preferably the same coefficients of thermal expansion. From the point of view of engineering, we use Si wafer substrates not only for the qubit chips but also for the interposers and package substrates because of the stable quality and the affordable cost. A 210-mm-square package substrate is currently obtainable from a commercially-available 300-mm Si wafer. In the near future, probably within a few years, 450-mm Si wafers will appear in the market and a 315-mm-square package substrate will be available for the QUIP packaging.

A single-chip QA factolizer comprising a 2 x 2 array of the multiplier units in Fig. 2(c) was laid out and then fabricated (Fig. 10). The unit consists of 12 qubits, 6 for functional operation and 6 for interconnection. The present version of the layout occupies an effective area of 495 μm × 510 μm per unit which is expected to increase to 515 μm × 530 μm per unit by adding the flip-chip bonding bumps for the MCM implementation, which suggests a possible integration of 35 × 35 units on a 19-mm-square chip. With these numbers, we formally expect factoring of a 350-bit integer by a QA machine consisting of 100 qubit chips on a 210-mm-square package substrate as illustrated in Fig. 11. Perhaps the number of bits to be factored, 350, is not very impressive but nevertheless the integration scale, exceeding one million qubits in total, is very challenging. Our future work will include more efficient designs of the QA Boolean logic circuits with the smaller number of qubits for constructing practical-scale QA machines for practical problems. There also remain a lot of issues to be considered, *e.g.*, a method of calibrating the individual qubits, arrangements of the fine tuning mechanisms such as the CCJJ SQUID and Iqp compensator,[11-14] and reduction of the number of cables between low and room temperatures.

## 6. Summary

We have designed a superconducting QA machine to perform prime factoring for analyzing RSA cryptosystem. The operation principle is the invertible operation of a multiplier composed of QA Boolean logic circuits, whose feasibility has been confirmed by circuit simulations with classical noise. We have proposed a new concept of special purpose QA machines, an ASAC architecture, which efficiently utilizes the limited hardware budget in superconducting integrated circuit technology in exchange for the restriction of the functionality. A proto-type single-chip version of a QA factorizer consisting of four unit cells was fabricated by using a Nb/AlO$_x$/Nb-junction superconducting integrated circuit technology. A practically large-scale ASAC QA factorizer is to be realized in a 2.5-dimensional QUIP packaging composed of qubit chips, active and bridge interposers, and a package substrate which are integrated by using a superconducting flip-chip bonding technology.

**Acknowledgment**

We are indebted to H. Iwata for the skillful assistance in the superconducting integrated circuit fabrication. We are also grateful to T. Endo, M. Hioki, K. Inomata, T. Katashita, S. Kohjiro, T. Nakagawa, T. Yamada, and H. Yamamori for valuable discussions. This paper is partly based on results obtained from a project commissioned by the New Energy and Industrial Technology Development Organization (NEDO), Japan.

*E-mail: masaaki.maezawa@aist.go.jp

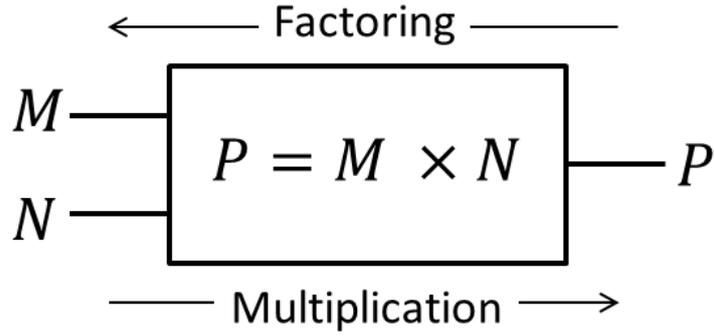

Fig. 1. Concept of factoring based on quantum annealing.

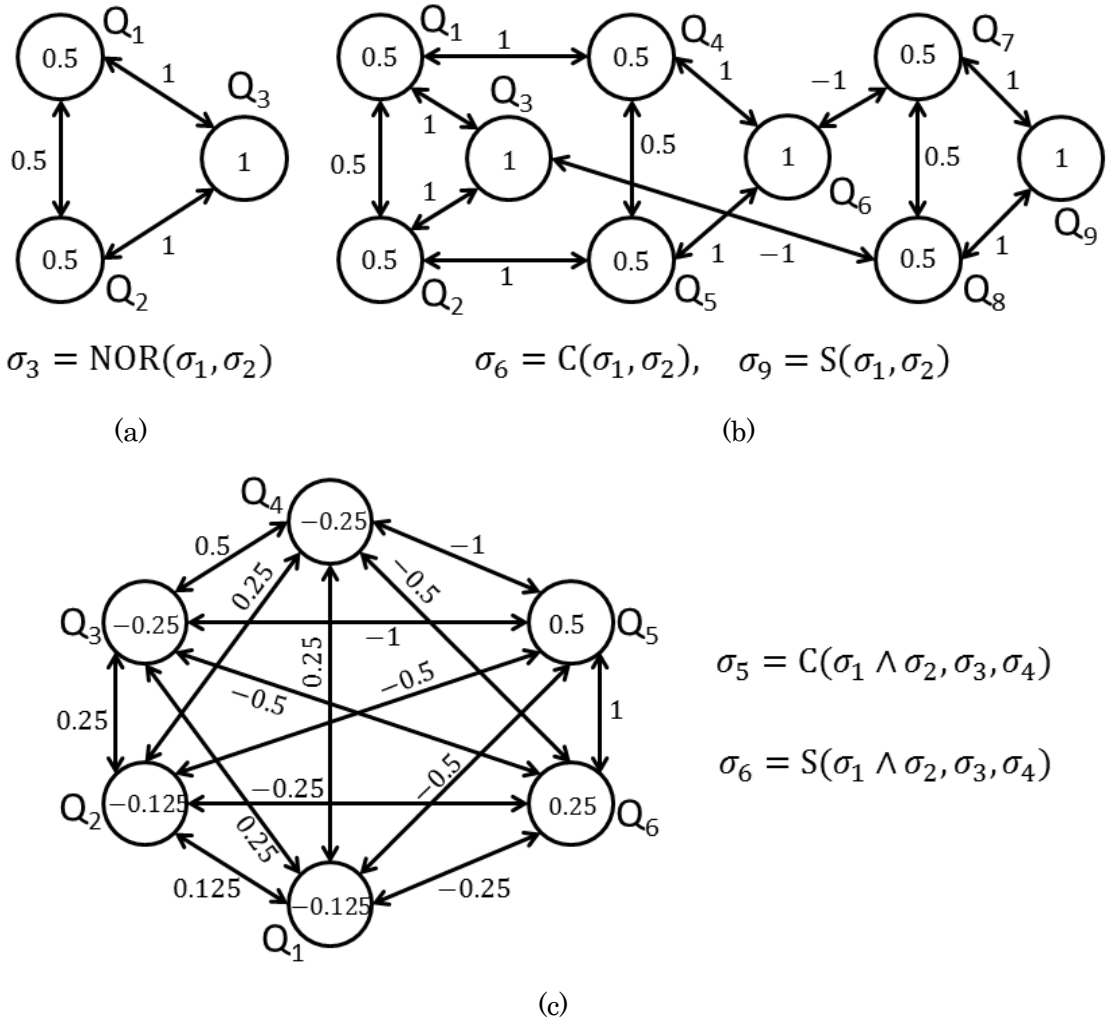

Fig. 2. Examples of QA implementations of Boolean logic circuits. (a) a NOR gate, (b) half adder and (c) multiplier unit. Circles denote qubits $Q_i$ with spin $\sigma_i$ and numbers inside are the coefficients $h_i$. Double sided arrows denote inter-qubit couplings between $Q_i$ and $Q_j$, and numbers aside are the coefficients $J_{ij}$.

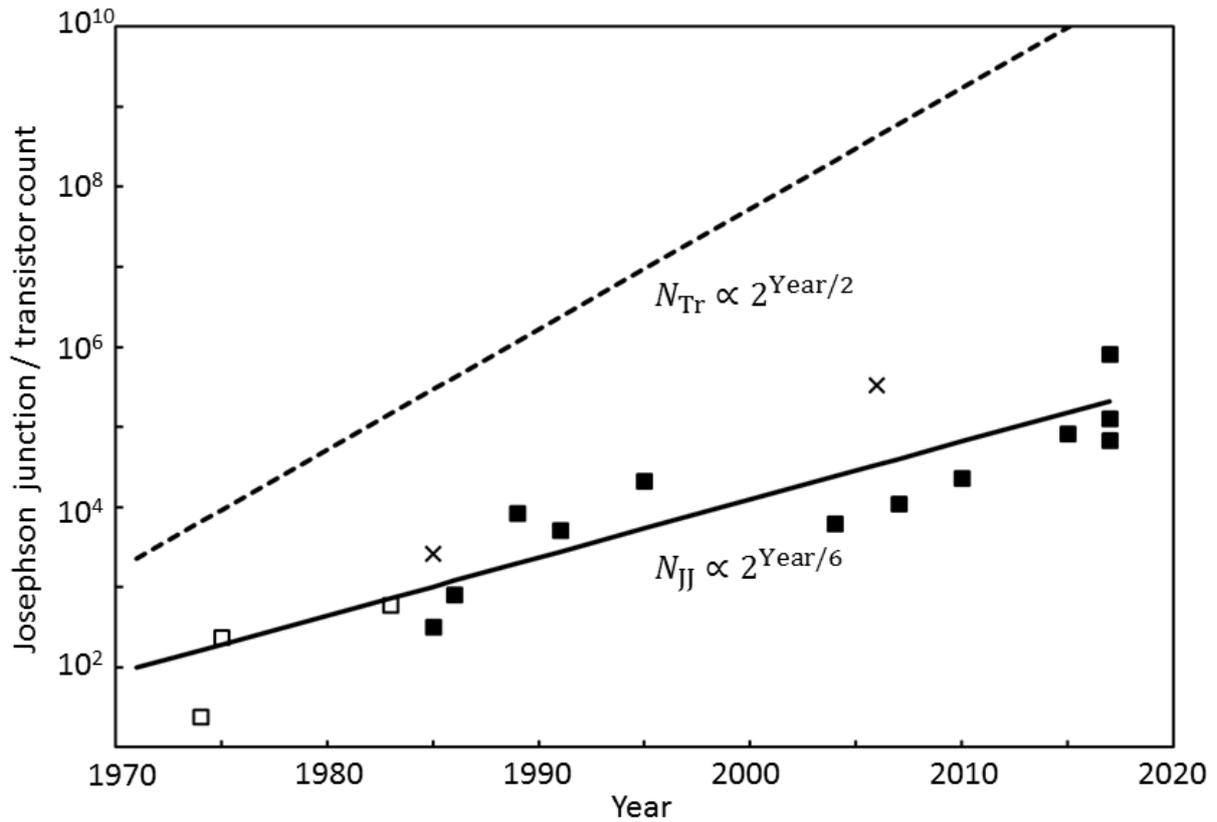

Fig. 3. Moore's law for Josephson junctions (solid line) and semiconductor transistors (dashed line). Open squares, solid squares, and crosses represent data reported for Pb-,[21] Nb-,[10,12,21,23-26,28,29] and NbN-[22,27] based Josephson junctions. The transistor count doubles every 2 years, whereas the Josephson junction count doubles every 6 years.

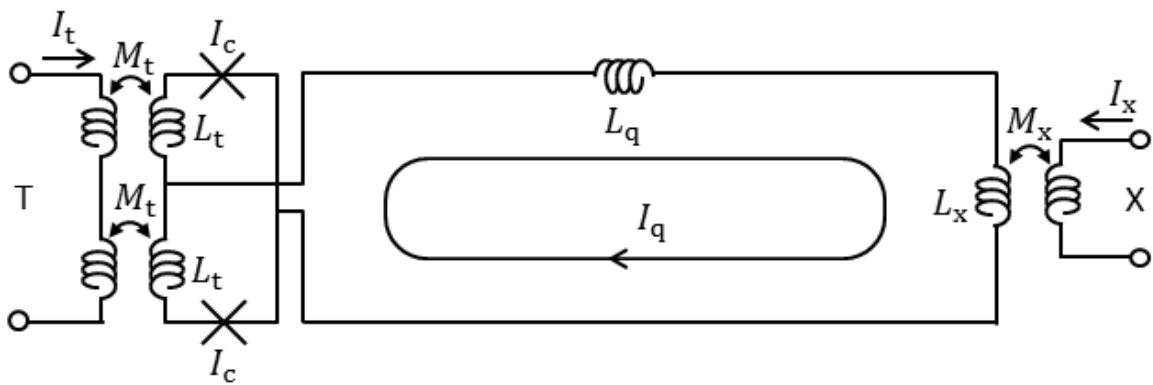

Fig. 4. A schematic of a flux qubit for QA factoring. A cross denotes a Josephson junction. The qubit-bias and transverse fields are induced by currents $I_x$ and $I_t$, respectively. The circuit parameters are $I_c = 4$ μA, $L_t = 5$ pH, $L_q = 250$ pH, $L_x = 10$ pH, $L_x = 10$ pH, $M_t = 2$ pH, and $M_x = 4$ pH.

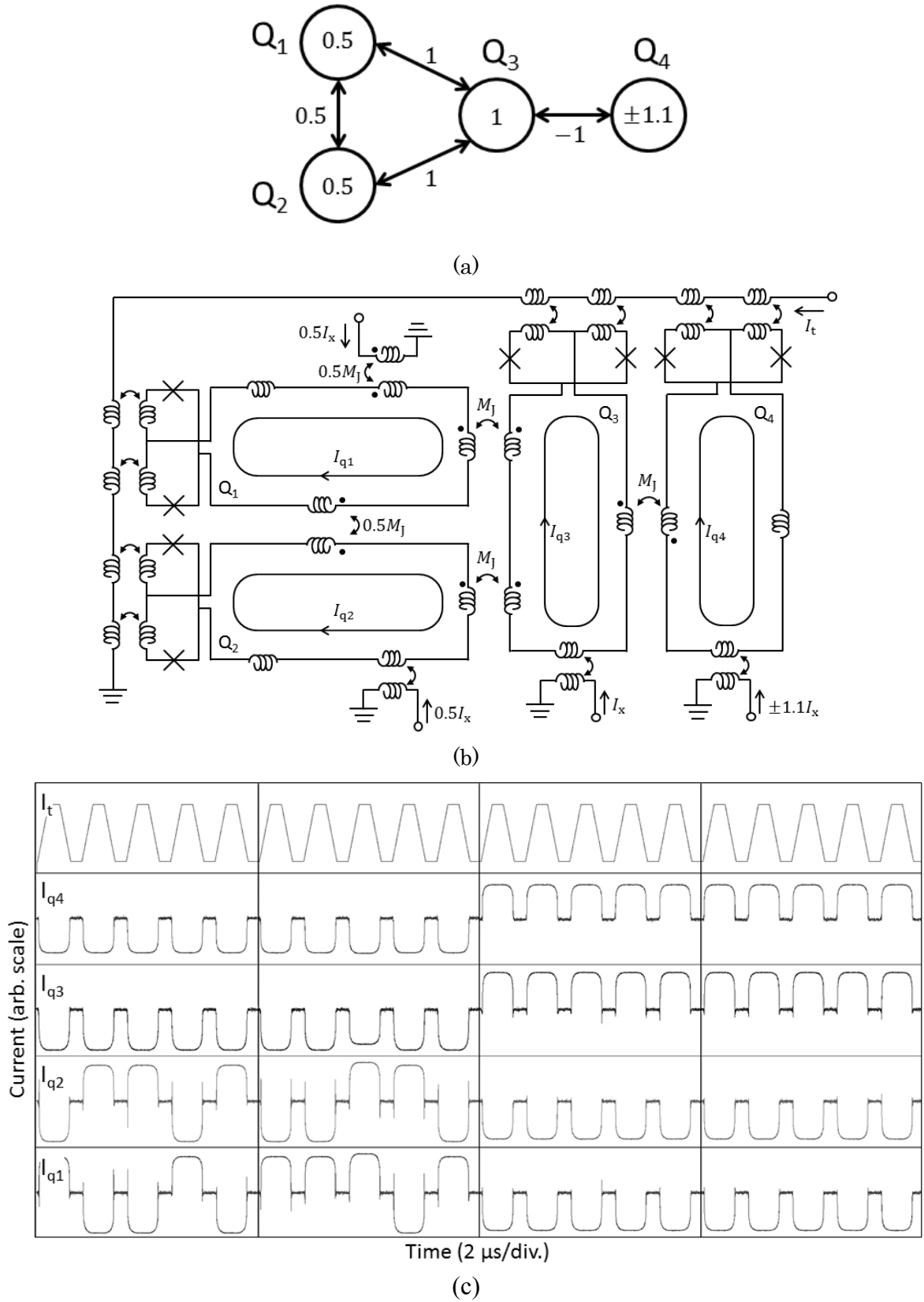

Fig. 5. Simulation of inverting QA NOR. (a) Circuit configuration in which a QA NOR consisting of $Q_1$, $Q_2$ and $Q_3$ has an extra qubit $Q_4$ for controlling the $Q_3$ state. (b) Schematic: $M_J$ = 8 pH and $I_x$ = 10.5 μA. (c) Waveforms of 20 shots of annealing operation.

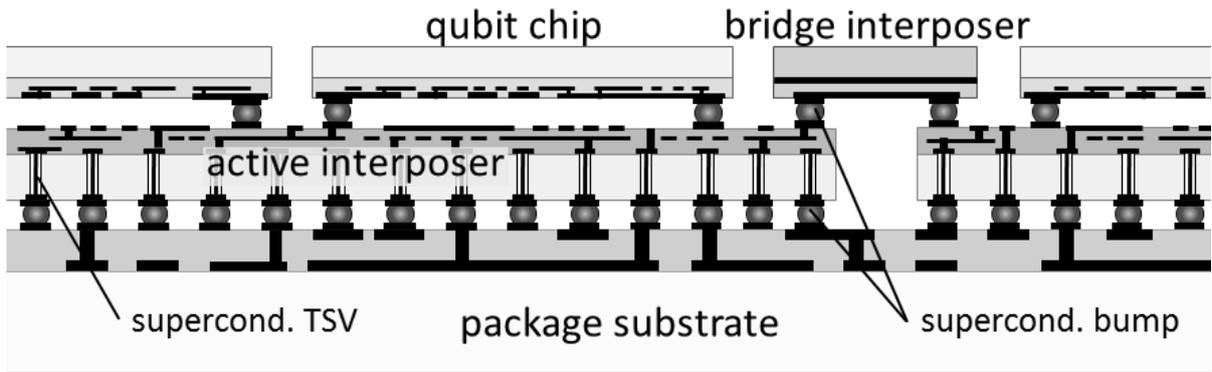

Fig. 6. QUbit-chip / Interposer / Package-substrate (QUIP) packaging for practical-scale QA machine.

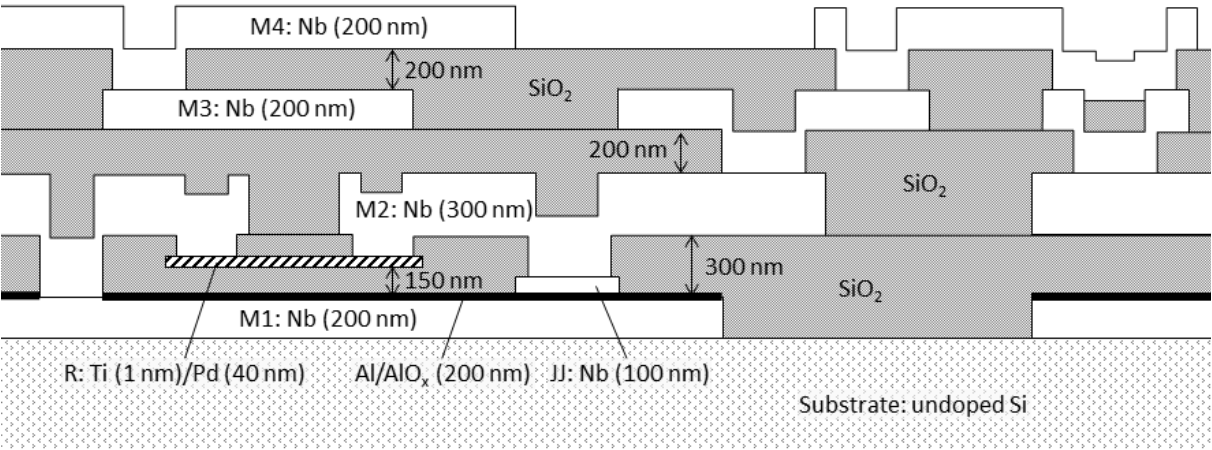

Fig. 7. A cross-sectional view of a qubit chip.

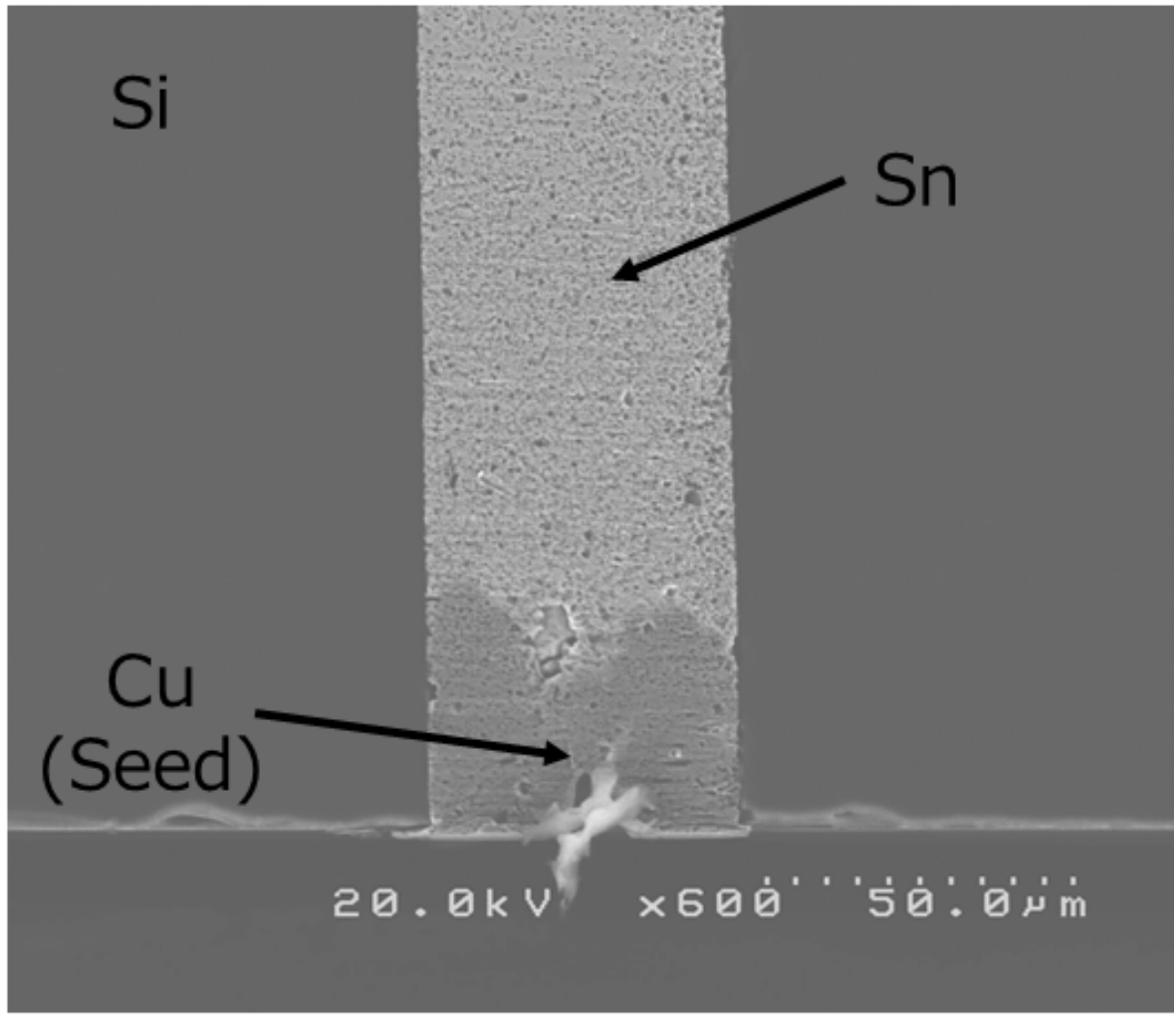

Fig. 8. A cross-sectional scanning electron microscope image of a superconducting through silicon via (TSV) made from tin with a nominal diameter of 50 μm.

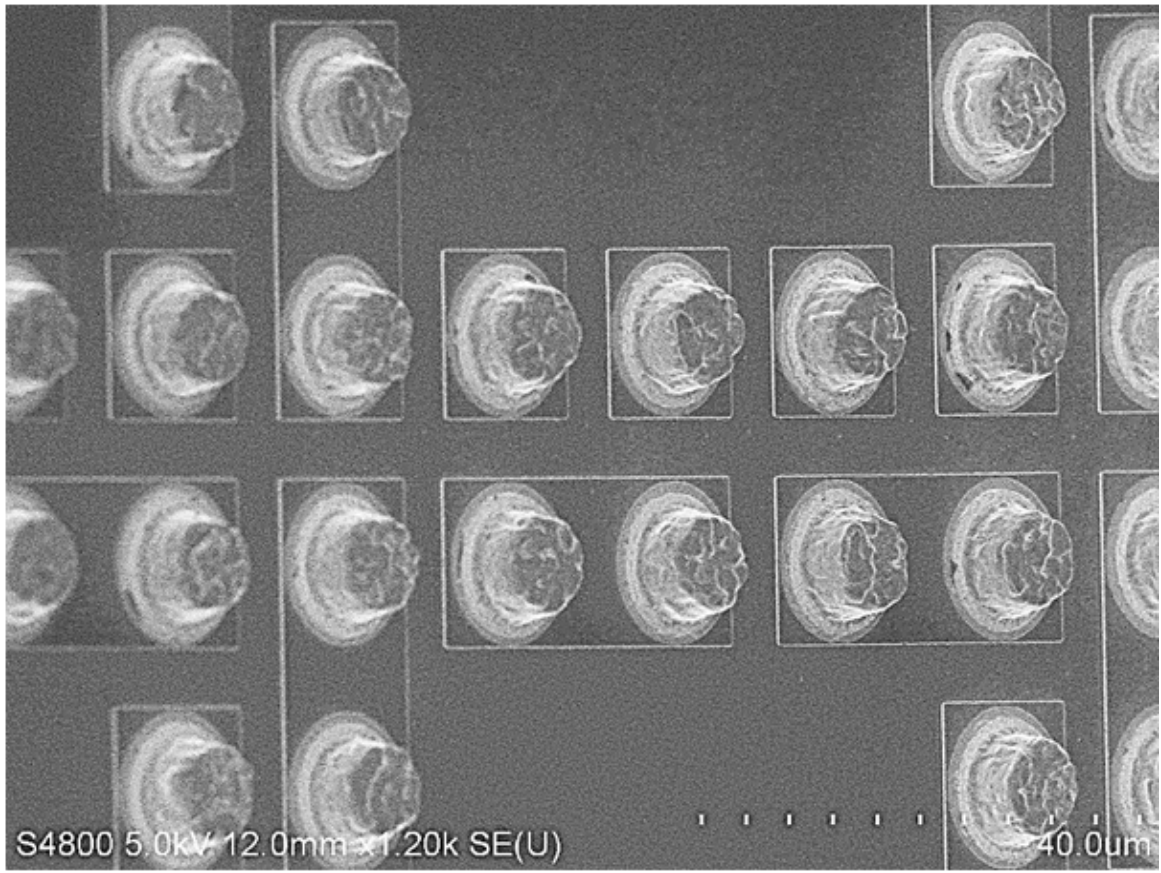

Fig. 9. A scanning electron microscope image of lead-alloy bumps. The nominal diameter and height of the bumps are 10 μm and 5 μm, respectively.

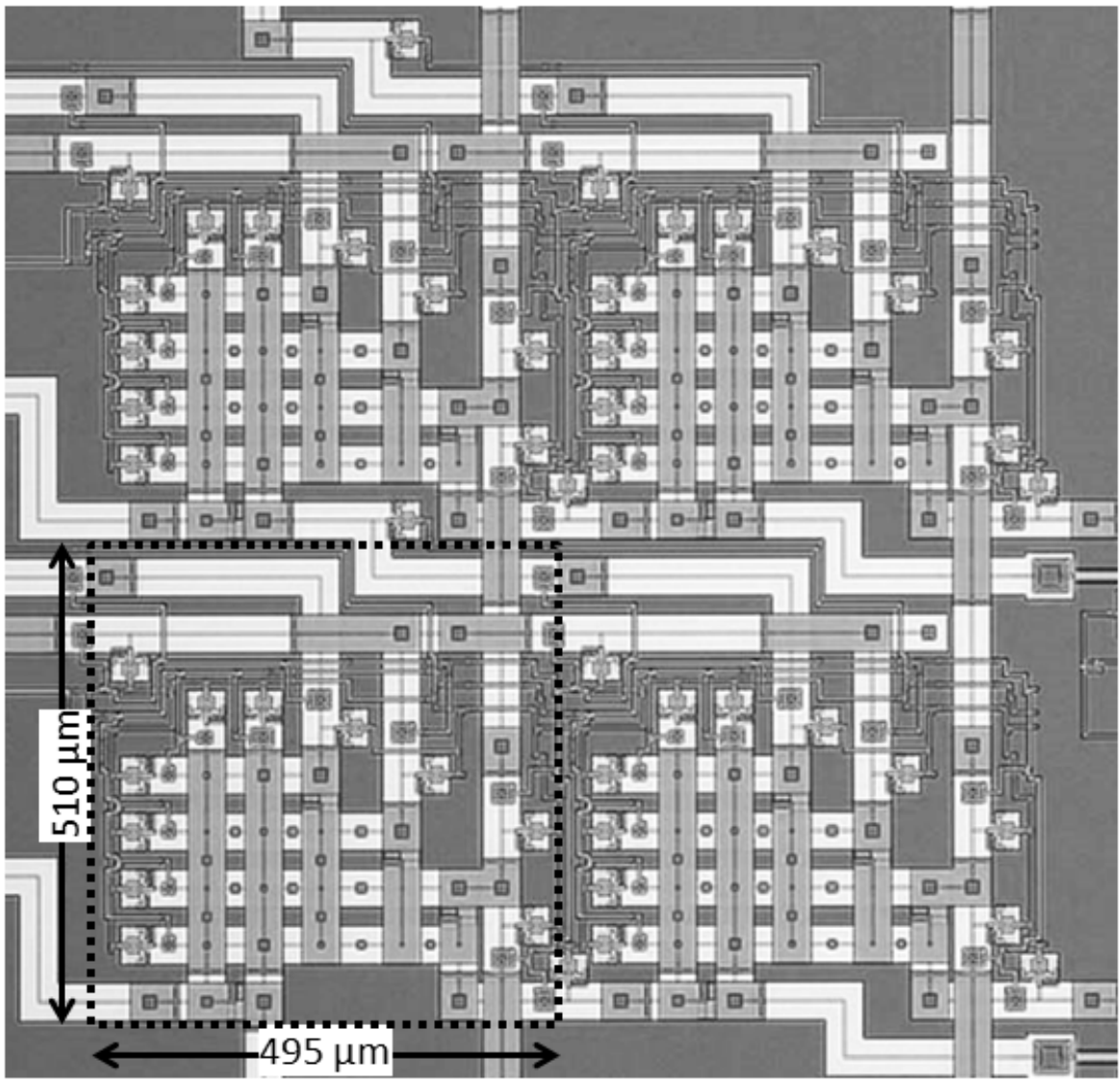

Fig. 10. A micrograph of a QA factorizer consisting of four unit cells. An effective area per unit, 495 μm × 510 μm, is denoted by a dotted box.

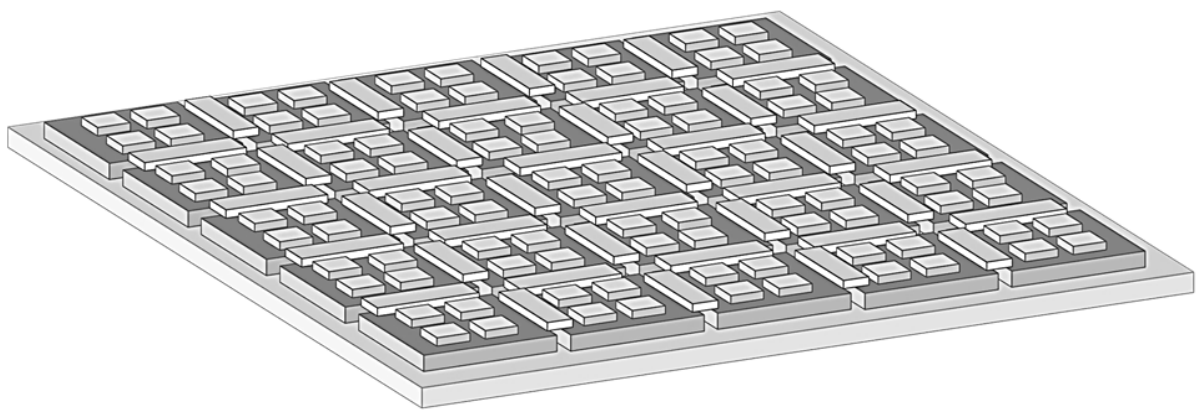

Fig. 11. A possible scheme toward integration of over 1 million qubits for 350-bit factoring.

Table I. Number of appearance of the combinations of the qubit states for inverting QA NOR simulation.

|  | $\sigma_3 = \sigma_4 = -1$ | $\sigma_3 = \sigma_4 = 1$ |
|---|---|---|
| $\sigma_1 = -1, \sigma_2 = -1$ | 0 | 100 |
| $\sigma_1 = -1, \sigma_2 = 1$ | 33 | 0 |
| $\sigma_1 = 1, \sigma_2 = -1$ | 47 | 0 |
| $\sigma_1 = 1, \sigma_2 = 1$ | 20 | 0 |